# The Vacuum and the Cosmological Constant Problem

Gerald E. Marsh

Argonne National Laboratory (Ret)
5433 East View Park
Chicago, IL 60615

E-mail: geraldemarsh63@yahoo.com

**Abstract.** It will be argued here that the cosmological constant problem exists because of the way the vacuum is defined in quantum field theory. It has been known for some time that for QFT to be gauge invariant certain terms—such as part of the vacuum polarization tensor—must be eliminated either explicitly or by some form of regularization followed by renormalization. It has recently been shown that lack of gauge invariance is a result of the way the vacuum is defined, and redefining the vacuum so that the theory is gauge invariant may also offer a solution to the cosmological constant problem.

**Introduction.**

The cosmological constant problem exists because of a number of key assumptions. These will be identified in this introduction. The second section of the paper discusses the issue of vacuum instability and negative energy states. Following this is a section on the relation between gauge invariance, Schwinger terms, and the definition of the vacuum. Given recent observational data from the Supernova Cosmology Project—showing an acceleration of the expansion of the universe at great distances, it is attractive to look at definitions of the vacuum that could lead to reasonable, but non-zero values of the cosmological constant. The redefinition of the vacuum proposed by Solomon [1] is one possible approach to achieving this goal. The final section looks at the origin of the difficulties in QFT and also serves as a summary.

Perhaps the best introduction to the cosmological constant problem is the review by Weinberg [2] published over twenty years ago. Although there have been a variety of approaches to the issue since then, the article can still serve to frame the problem, at least in the sense that it is to be addressed here. To avoid confusion, Weinberg's notation will be used in this introduction.

The Einstein field equations, including the cosmological constant $\lambda$, are

$$R_{\mu\nu} - \frac{1}{2} g_{\mu\nu} R - \lambda\, g_{\mu\nu} = -8\pi G\, T_{\mu\nu}. \tag{1}$$

The right hand side of the equation contains the energy-momentum tensor $T_{\mu\nu}$, and it is here that the first assumption that leads to the cosmological constant problem is made. It is that the vacuum has a non-zero energy density. If such a vacuum energy density exists, Lorentz invariance requires that it have the form*

$$\langle T_{\mu\nu} \rangle_{Vac} = -\langle \rho \rangle\, g_{\mu\nu}. \tag{2}$$

This allows one to define an effective cosmological constant and a total effective vacuum energy density

$$\begin{aligned} \lambda_{eff} &= \lambda + 8\pi G \langle \rho \rangle \\ \rho_V &= \langle \rho \rangle + \frac{\lambda}{8\pi G} = \frac{\lambda_{eff}}{8\pi G}. \end{aligned} \tag{3}$$

* The signature of the metric is +2.

Observationally, it is known that the total effective vacuum energy density must be comparable to about $\rho_V \sim 10^{-29}\ \frac{g}{cm^3}$.

The second assumption that leads to the cosmological constant problem is that the vacuum energy density is due to the zero-point energy of a quantized field or fields. The concept of the zero-point energy originated with the quantization of the simple harmonic oscillations of a particle with non-vanishing mass. This energy is present for each energy level including the ground state. In QFT, at a given instant of time, the field is defined at each point in space—essentially an independent harmonic oscillation at each point with amplitude and phase depending on the initial conditions. By specifying an equation of motion depending on the amplitude of the field and its partial derivative with respect to time, one couples these otherwise independent oscillations. In the Klein-Gordon equation, for example, this coupling is achieved by the presence of the Laplacian.

The zero point energy is also present in the relativistic QFT of the neutral Klein-Gordon field and the electromagnetic field. Of course the zero-point energy of the Klein-Gordon field is infinite, but when used to formulate the cosmological constant problem it is generally summed up to a cutoff that determines the magnitude of the vacuum energy density. The way this is done is to count the number of normal modes between $\nu$ and $\nu + d\nu$, multiply by the zero-point energy in each mode, and integrate up to the chosen cutoff. There is some question as to whether it makes sense to count the normal modes for a massive particle when addressing the energy density of the vacuum where no particles are present. The usual justification for this is somewhat confused in the literature, but is generally based on identifying the zero-point energy with vacuum fluctuations.

The number of normal modes between $\nu$ and $\nu + d\nu$ is given by

$$dZ = \frac{4\pi V}{c^3}\nu^2\,d\nu. \tag{4}$$

Transforming to wave number, setting the volume equal to unity, and using units where $\hbar = c = 1$, and integrating up to a wave number cutoff $\Lambda >> 1$ results in a vacuum energy density of

$$\langle \rho \rangle \; = \; \int_0^{\Lambda} \frac{1}{2}\sqrt{k^2 + m^2}\,\frac{4\pi\, k^2}{(2\pi)^3}\,\mathrm{d}k \;\; - \frac{\Lambda^4}{16\,\pi^2}. \tag{5}$$

As a relativistic wave equation, the Klein-Gordon equation has an energy spectrum that includes both positive and negative energies, corresponding to either sign for the radical in Eq. (5). Note that if the integral were performed for both signs and the results added, the vacuum energy density would vanish. In QFT, zero-point energies are inherent in the canonical field quantization method because the ordering of operators in the Hamiltonian is not fixed.

In Eq. (5), the vacuum energy density up to a cutoff is calculated for the cosmological constant problem by using only the positive sign for the radical. The reason for this goes back to Dirac's redefinition of the vacuum where electrons are assumed to fill the negative-energy states, and they and their infinite vacuum charge density are presumed to be unobservable—as are the negative-energy state zero-point energies. In this way the instability of the vacuum caused by the availability of negative energy states is eliminated. The normal ordering procedure of canonical QFT, in addition to eliminating the zero-point energies, also has the virtue of eliminating the infinite vacuum charge density of the Dirac vacuum. QFT does not, however, eliminate the problem of negative energies as will be seen below.

The value of wave number cutoff to be chosen in Eq. (5) depends on one's view of Einstein's theory of gravitation. Fields are generally defined in the context of a background metric for space-time—the Minkowski metric. General relativity has to do with the geometry of space-time itself, and while there is no *a priori* reason (or experimental evidence) that the gravitational field need be quantized, it is generally believed that Einstein's theory of gravitation will no longer hold at very small distances, and in particular for distances comparable or smaller than the Planck length $(G\hbar/c^3)^{1/2}$. At this length scale, quantum fluctuations are thought to be the dominant influence on the local space-time geometry. For this reason, the Planck length is generally chosen to define the cutoff. Such a cutoff is really an expression of our ignorance of the vacuum and the structure (if there is one) of space-time itself at these distances. Unfortunately, no experimental data exists.

Evaluating the right hand side of Eq. (5) with Weinberg's choice of $\Lambda = (8\pi G)^{-1/2}$, (the units are again such that $\hbar = c = 1$) results in a value for the vacuum energy density of

$$\langle \rho \rangle = 2 \text{ X } 10^{89} \frac{\text{g}}{\text{cm}^3}. \tag{6}$$

Comparing this with the observational value of $\rho_V \sim 10^{-29}$ g/cm$^3$ tells us that the two terms in the effective vacuum energy density of Eq. (3) must cancel to some 118 decimal places. This fine tuning *is* the cosmological constant problem as it is currently understood.

The Casimir effect, which in QFT is thought to be due to the presence of vacuum fluctuations, is used to argue that the zero-point energies are real. That is, zero-point energies and quantum fluctuations are identified. A key reference for the Casimir effect is the work of Plunien, Müller, and Greiner [3]. Vacuum fluctuations—which represent the terms of the perturbation series for the vacuum expectation value $<0|S|0>$ of the $S$-matrix—are generally ignored since they can only produce an overall phase factor. However, in the presence of external sources or boundaries these fluctuations can no longer be so casually dismissed. Although the vacuum state must be invariant under rotations and translations—and therefore must have zero momentum, angular momentum, and energy—the presence of external sources or boundaries (as in the Casimir effect) breaks these fundamental symmetries, allowing vacuum fluctuations to have effects that are observable.

Moreover, if the energy of the vacuum is defined as the difference of zero-point energies in the presence of boundaries $\partial\Gamma$ of a region $\Gamma$ and without boundaries [3], the vacuum energy can be *negative*; that is,

$$E_{vac}[\partial\Gamma] = E_0[\partial\Gamma] - E_0[0], \tag{7}$$

where $E_0[\partial\Gamma]$ is the zero-point energy in the presence of boundaries and $E_0[0]$ is the zero-point energy in their absence.

Because of the absence of boundaries or external sources, performing the kind of calculation given in Eq. (5) has raised questions in the literature with regard to the legitimacy of the approach of simply summing free field modes. See, for example,

Lamoreaux [4] where it is argued that we don't learn much about the properties of the vacuum of free space through the study of Casimir and related zero-point energy effects.

In the absence of interactions as well as boundaries, one could make the argument that there appears to be no reason not to allow the vacuum to have a negative energy spectrum so that both signs of the energy in evaluating Equation (5) should be used, thereby eliminating the cosmological constant problem. In the presence of interactions the issue becomes more complicated, but—as will seen below—the availability of a negative energy spectrum is one way to restore gauge invariance.

**Vacuum Instability and Negative Energy States**

At this point it might be useful to examine the origin of the concern about vacuum instability in the presence of negative energy states. Consider a free electron in a positive energy state $E$ subjected to a periodic perturbation of frequency $\omega$. It is generally agreed that there is a non-vanishing probability for the electron to make a transition to a state of energy $E + \hbar\omega$ or $E - \hbar\omega$. If $\hbar\omega > E + mc^2$, it is argued that the transition to $E - \hbar\omega$ would be to a state of negative energy. Having set the stage, the argument continues by considering a bound state electron in a hydrogen atom. Since the electron is coupled to the electromagnetic field, such a bound state would rapidly make a radiative transition to a negative energy state, with the result that the hydrogen atom would have no stable existence. Worse yet, since the spectrum has no lower bound, there would be no limit to the radiated energy.

There is an objection to the above argument that shows that such transitions could well be forbidden. Under the usual sign convention, quantum states of positive energy evolve in time as $e^{-i\omega t}$. A state of negative energy, $-E$, then evolves as $e^{+i\omega t}$, which corresponds to the transformation $t \rightarrow -t$. But there are two possibilities for a time reversal operator: it can be unitary or anti-unitary—where $t \rightarrow -t$ and one also takes the complex conjugate of states and complex numbers. Under a unitary transformation, used for all other discrete and continuous symmetries, the time-reversed state corresponds to a state of negative energy $-E$. It was Eugene Wigner who introduced the anti-unitary time reversal operator so as to eliminate negative energies.

There are now two arguments that can be given against the transition from a bound state (as in the above argument) to a negative energy state under the periodic perturbation. First, if the Hamiltonian $H$ governing the transition is to be *CPT* invariant—as it must if it is to be an acceptable quantum electrodynamics Hamiltonian—it must satisfy $CPT\, H(x)\, [CPT]^{-1} = H(-x)$. This will only be the case if the operator *CPT* is anti-unitary, a consequence of the complex conjugation implicit in the time reversal operator *T*. But since the transformation from $e^{-i\omega t} \to e^{+i\omega t}$ corresponds to simply $t \to -t$, we have instead $CPT\, H(x)\, [CPT]^{-1} = -H(-x)$, giving the negative energy state. Thus, the Hamiltonian of the transition can not satisfy *CPT* invariance.

The second argument has to do with the fact that if an electron in an external field obeys the (quantized) Dirac equation, one cannot rule out the negative energy solutions needed to make up a complete set of wave functions. But a wave function representing a negative energy state can only be non-zero if it has charge $+e$. [5] This means the argument given above for the transition of an electron to a negative energy state will violate charge conservation.

Thus, a radiative transition of an electron to a negative energy state either violates *CPT* invariance or the conservation of charge. So it would appear that such transitions are effectively forbidden. This means there is no reason not to allow negative energies to be summed over in Eq. (5), yielding a vanishing vacuum energy density. The attractiveness of Solomon's approach to redefining the vacuum, to be described below, is that it allows for incomplete cancellation of the zero-point energies leading to a small, but not vanishing vacuum energy density.

**QFT and Gauge Invariance.**

The issue of the gauge invariance of QFT has been dealt with in a variety of ways over the years (an extensive discussion is contained in [7]). In essence, the standard vacuum of QFT is only gauge invariant if non-gauge invariant terms are removed. There are two general approaches to the problem: the first is to simply ignore such terms as being physically untenable and remove them so as to maintain gauge invarianc; and the second is to use various regularization techniques to cancel the terms. Pauli-Villars regularization [6] is discussed below.

Returning to Eq. (5), another problem with this equation is that the introduction of a momentum-space cutoff destroys translational invariance and may make it difficult to maintain gauge and Lorentz invariance. If, instead of introducing a cutoff, an attempt is made to deal with this infinite integral by Pauli-Villars regularization, the result is the introduction of negative masses. This can be seen as follows. It can be guaranteed that the vacuum expectation value of the energy momentum tensor is Lorentz invariant and hence proportional to $\eta^{\mu\nu}$ if we use the relativistic formulation

$$\langle 0 | T^{\mu\nu} | 0 \rangle = \frac{1}{2} \int_0^\infty \frac{dp^3}{(2\pi)^3} \frac{p^\mu p^\nu}{p^0} . \tag{8}$$

Here $p^0 = E(\boldsymbol{p}) = (\boldsymbol{p}^2 + m^2)^{1/2}$. The domain of integration can be transformed to spherical coordinates in a space of any dimension by use of the formula

$$\prod_{i=1}^{d} dp^i = p^{d-1} dk \prod_{i=1}^{d-1} \sin^{i-1}\theta_i \, d\theta_i , \tag{9}$$

resulting in

$$\langle 0 | T^{00} | 0 \rangle = \frac{1}{2} \int_0^\infty \frac{4\pi \, p^2 dp}{2\,(2\pi)^3} \sqrt{p^2 + m^2} . \tag{10}$$

The Pauli-Villars regulator masses $m_i$, where $m_1 = m$, and associated coefficients $c_i$, where $c_1 = 1$, are then introduced as follows:

$$\langle 0 | T^{00} | 0 \rangle = \frac{1}{2} \int_0^\infty \frac{4\pi \, p^2 dp}{2\,(2\pi)^3} \sum_i c_i \, p \sqrt{1 + \frac{m_i^2}{p^2}} . \tag{11}$$

The number of regulator masses needed depends in general on the integral. Since we are interested in the convergence of the integral at the upper limit where $m_i^2/p^2 < 1$, the radical can be expanded in a series to yield

$$\langle 0 | T^{00} | 0 \rangle = \frac{1}{2} \int^\infty \frac{4\pi \, p^2 dp}{2\,(2\pi)^3} \sum_i c_i \left( p + \frac{1}{2} \frac{m_i^2}{p} - \frac{m_i^4}{8p^3} + \ldots \right) . \tag{12}$$

This integral will converge at the upper limit provided the following relations are satisfied as $m_i \to \infty$:

$$1 + \sum_{j=2}^{n} c_j \; = \; 0 \qquad m^2 + \sum_{j=2}^{n} c_j \, m_j^2 \; = \; 0 \qquad m^4 + \sum_{j=2}^{n} c_j \, m_j^4 \; = \; 0 \qquad m_j \to \infty. \tag{13}$$

Moving $p^2$ in the numerator of Eq. (12) into the series expansion shows that Eqs. (13) will be satisfied if $n = 3$.

Because some of the coefficients must be negative, this procedure introduces negative energies in the form of negative masses that are allowed to become infinite at the end of the calculation. The advantage, at least for QFT, is that gauge and Lorentz invariance are preserved.

If one works in Euclidean space by first performing a Wick rotation, one is left with essentially the same problem: regulator fields for scalar fields obey Fermi statistics, and those for spinor fields obey Bose statistics. This violation of the spin-statistics theorem means that the Hamiltonian cannot be a positive definite operator, again implying the existence of negative energy states.

Solomon [7] has argued that for QFT to be gauge invariant the Schwinger term must vanish, and Schwinger [8] long ago showed that for this to be the case the vacuum state cannot be the state with the lowest free field energy. The existence of non-zero Schwinger terms also impacts Lorentz invariance. Lev [9] has shown that if the Schwinger terms do not vanish the usual current operator $\hat{J}^{\mu}(x)$, where $\mu = 0, 1, 2, 3$ and $x$ is a point in Minkowski space, is not Lorentz invariant.

To begin with, however, it is important to understand the details of Schwinger's argument. Using Solomon's notation, the Schwinger term is given by

$$\mathrm{ST}(\vec{y},\vec{x}) \; = \; \left[\hat{\rho}(\vec{y}),\, \hat{\vec{J}}(\vec{x})\right]. \tag{14}$$

Taking the divergence of the Schwinger term and using the relation (see [7])

$$\mathrm{i}\left[\hat{\mathrm{H}}_0,\, \hat{\rho}(\vec{x})\right] \; = \; - \quad \hat{\vec{J}}(\vec{x}), \tag{15}$$

where $\hat{\mathrm{H}}_0$ is the free-field Hamiltonian when the electromagnetic 4-potential vanishes, results in

$$_{\vec{x}} \left[\hat{\rho}(\vec{y}),\, \hat{\vec{J}}(\vec{x})\right] \; = \; \left[\hat{\rho}(\vec{y}), \quad \hat{\vec{J}}(\vec{x})\right] \; = \; -\mathrm{i}\left[\hat{\rho}(\vec{y}),\, \left[\hat{\mathrm{H}}_0,\, \hat{\rho}(\vec{x})\right]\right]. \tag{16}$$

Expanding the commutator on the right hand side of Eq. (16) yields the vacuum expectation value

$$i \quad {}_{\vec{x}} \left\langle 0 \left| \left[ \hat{\rho}(\vec{y}),\, \hat{\vec{J}}(\vec{x}) \right] \right| 0 \right\rangle \;=\; -\left\langle 0 \right| \hat{\mathrm{H}}_0\, \hat{\rho}(\vec{x})\hat{\rho}(\vec{y}) \left| 0 \right\rangle +$$
$$\left\langle 0 \right|\; \hat{\rho}(\vec{x})\hat{\mathrm{H}}_0\hat{\rho}(\vec{y}) \left| 0 \right\rangle + \left\langle 0 \right|\; \hat{\rho}(\vec{y})\hat{\mathrm{H}}_0\hat{\rho}(\vec{x}) \left| 0 \right\rangle - \left\langle 0 \right|\; \hat{\rho}(\vec{y})\hat{\rho}(\vec{x})\hat{\mathrm{H}}_0 \left| 0 \right\rangle. \tag{17}$$

It is here that one makes the assumption that the vacuum is the lowest energy state. This done by writing $\hat{\mathrm{H}}_0|0> = <0|\hat{\mathrm{H}}_0 = 0$. As a result, Eq. (17) may be written as

$$i \quad {}_{\vec{x}} \left\langle 0 \left| \left[ \hat{\rho}(\vec{y}),\, \hat{\vec{J}}(\vec{x}) \right] \right| 0 \right\rangle \;=\; \left\langle 0 \right|\; \hat{\rho}(\vec{x})\hat{\mathrm{H}}_0\hat{\rho}(\vec{y}) \left| 0 \right\rangle +$$
$$\left\langle 0 \right|\; \hat{\rho}(\vec{y})\hat{\mathrm{H}}_0\hat{\rho}(\vec{x}) \left| 0 \right\rangle. \tag{18}$$

Multiply both sides of the last equation by $f(x)f(y)$ and integrate over $x$ and $y$. The right hand side of Eq. (18) becomes

$$\int d\vec{x}\; d\vec{y}\; \left\{ \left\langle 0 \right|\; f(\vec{x})\hat{\rho}(\vec{x})\hat{\mathrm{H}}_0 f(\vec{y})\hat{\rho}(\vec{y}) \left| 0 \right\rangle + \left\langle 0 \right|\; f(\vec{y})\hat{\rho}(\vec{y})\hat{\mathrm{H}}_0 f(\vec{x})\hat{\rho}(\vec{x}) \left| 0 \right\rangle \right\}. \tag{19}$$

If Schwinger's "arbitrary linear functional of the charge density" is defined as

$$F \;=\; \int f(\vec{x})\hat{\rho}(\vec{x})d\vec{x} \;=\; \int f(\vec{y})\hat{\rho}(\vec{y})d\vec{y}\;\;, \tag{20}$$

the right hand side of Eq. (18) becomes

$$2\left\langle 0 \right| F\hat{\mathrm{H}}_0 F \left| 0 \right\rangle \;=\; 2\sum_{m,n} \langle 0|\, F\, |m\rangle \left\langle m \right| \hat{\mathrm{H}}_0 \left| n \right\rangle \langle n|\, F\, |0\rangle \;=$$
$$2\sum_n E_n \langle 0|\, F\, |n\rangle\langle n|\, F\, |0\rangle \;=\; 2\sum_n E_n |\, \langle 0|\, F\, |n\rangle\, |^2 > 0. \tag{21}$$

The left hand side of Eq. (21)—essentially the form used by Schwinger—is here expanded to explicitly show the non-vanishing matrix elements between the vacuum and the other states of necessarily positive energy. This shows that if the vacuum is assumed to be the lowest energy state, the Schwinger term cannot vanish, and the theory is not gauge invariant. Solomon also shows the converse, that if the Schwinger term vanishes, then the vacuum is not the lowest energy state and the theory *is* gauge invariant.

For the sake of completeness, it is readily shown that the left side of Eq. (18) becomes

$$i\int \quad {}_{\vec{x}} \left\langle 0 \left| \left[ \hat{\rho}(\vec{y}),\, \hat{\vec{J}}(\vec{x}) \right] \right| 0 \right\rangle\, f(\vec{x})f(\vec{y})\mathrm{d}\vec{x}\mathrm{d}\vec{y} \;=\; i\, \langle 0\, |\, [\partial_t F,\, F]\, |0\rangle, \tag{22}$$

so that combining Eqs. (21) and (22) yields a somewhat more explicit form of the result given by Schwinger,

$$i\, \langle 0\, |\, [\partial_t F,\, F]\, |0\rangle \;=\; 2\sum_n E_n |\, \langle 0|\, F\, |n\rangle\, |^2 > 0. \tag{23}$$

**Solomon's redefinition of the vacuum**

In QFT, invariant perturbation theory [10] the leads to the Dyson chronological operator $P(H_I(\mathbf{x}_1) \ . \ . \ . \ H_I(\mathbf{x}_n))$, where $H_I$ is the interaction Hamiltonian. The adjective "invariant" refers to the fact that time ordering in the Dyson series is Lorentz invariant if the $H_I(\mathbf{x}_i)$ all commute at space-like separations. The chronological product can be expressed in a form where the virtual processes are explicitly represented; that is, as a decomposition into normal products, where a Feynman graph can be used to represent each of the normal products. The vacuum state is empty (although vacuum fluctuations exist) and is the state of lowest energy. The theory is not, however, gauge invariant and a process of regularization and renormalization is used to make it so.

The virtue of the vacuum state to be defined in this section is that it allows QFT to be mathematically consistent in the sense that it becomes gauge invariant without the need for regularization. Because the theory is gauge invariant, the Schwinger term vanishes. The reason for this is that the definition of the vacuum is such that the usual vacuum state, |0>, is no longer the state of minimum energy, and there exist states with negative energy. In the context of the cosmological constant problem, this means that even if one believes the summation of zero-point energies given in Eq. (5) is legitimate, it must be extended to the unoccupied negative energy states, leading to a significant cancellation in the summation.

Redefining the vacuum state may provide a more elegant means of resolving both the gauge invariance difficulties of QFT and the cosmological constant problem. The specific definition of the vacuum state given below serves, at a minimum, as an heuristic example of this approach.

Because of its intuitive nature, hole theory will be used to introduce Solomon's redefinition of the vacuum in quantum electrodynamics. He has also implemented this redefinition in the context of QFT, and that will also be described here. It may therefore be useful to recall how the transition to QFT is made.

It terms of positive and negative energy electrons, second quantization gives the Hamiltonian for a free electron

$$H_0 = \sum_{ps} E_p\left(a^\dagger_{ps}a_{ps} - b^\dagger_{ps}b_{ps}\right), \tag{24}$$

where $a^\dagger_{ps}$ is the creation operator for positive energy electrons and $b^\dagger_{ps}$ creates negative energy electrons. In terms of the number operator *N*,

$$H_0 = \sum_{ps} E_p\left[N^\dagger_{ps} - N^-_{ps}\right], \tag{25}$$

where $E_p > 0$ and $N^-_{ps}$ is the number of negative energy electrons.

Dirac's hole theory rescales the energy so that

$$H_0' = \sum_{ps} E_p\left[N^\dagger_{ps} + \left(1 - N^-_{ps}\right)\right]. \tag{26}$$

If $N^-$ (for a given *p* and *s*) vanishes—i.e., when a negative-energy electron is missing, then *N*+ increases by one and we see a positive energy electron and a hole in the negative energy continuum, which is interpreted as a positron.

For the hole theory approach to dealing with negative energies to work, it is essential that the particles filling the negative-energy states obey the Pauli exclusion principle. The technique would not work for bosons associated with the Klein-Gordon equation. While the particles filling the negative-energy states must not produce an electric field or contribute to the total charge, energy, or momentum, they nevertheless must respond to an external field.

The transition to QFT is accomplished by setting

$$a_{ps} = c_{ps} \quad \text{and} \quad b_{ps} = d^\dagger_{ps}, \tag{27}$$

so that destroying a negative energy electron is equivalent to creating a positron. The Dirac negative-energy sea vanishes since electrons and positrons are treated as separate entities. Provided the *b*'s satisfy the anti-commutation relations

$$\left\{b^\dagger_{ps}, b_{p's'}\right\} = \delta_{ss'}\delta_{pp'}, \tag{28}$$

we have that

$$N^{positrons}_{ps} = 1 - N^-_{ps} = 1 - b^\dagger_{ps}b_{ps} = b_{ps}b^\dagger_{ps} = d^\dagger_{ps}d_{ps}$$

$$N^{electrons}_{ps} = N^\dagger_{ps} = a^\dagger_{ps}a_{ps} = c^\dagger_{ps}c_{ps}. \tag{29}$$

As a result,

$$N_{ps}^{positrons}|\,0\,\rangle \;=\; N_{ps}^{electrons}|\,0\,\rangle \;=\; 0, \tag{30}$$

since |0> contains neither positrons or electrons. The free-electron Hamiltonian in terms of the number operator, now sums over both electrons and positrons and is given by

$$\boldsymbol{H}_0 \;=\; \sum_{ps} E_p\left[N_{ps}^{positrons} + N_{ps}^{electrons}\right] \tag{31}$$

Thus, one can readily move between hole theory and QFT although, as Solomon has shown [11], one sometimes obtains differing results because of the way the vacua are defined.

With reference to Fig. 1, Solomon defines a state vector $|0, \Delta E_W>$ as the state where a band of negative energy states extending from $-m$ to $-(m + \Delta E_W)$ is occupied by a single particle (the exclusion principle holds); all other single particle states are unoccupied.

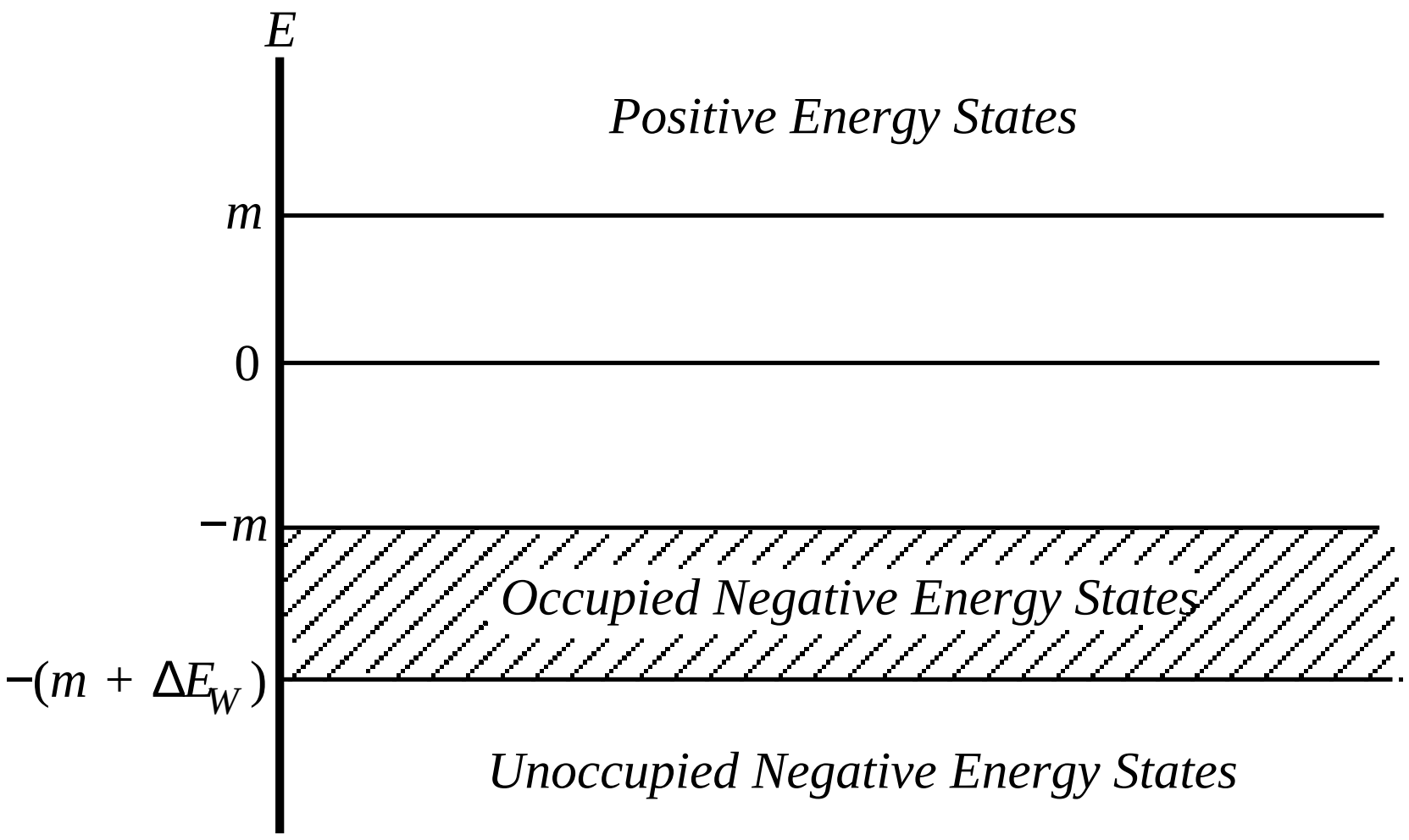

Figure 1. The state vector $|0, \Delta E_W>$. Only the band of negative energy states extending from $-m$ to $-(m + \Delta E_W)$ is occupied.

The vacuum state is defined by Solomon as $|0, \Delta E_W \rightarrow \infty>$, where it is important to understand that the limit $\Delta E_W \rightarrow \infty$ means that $\Delta E_W$ goes to an arbitrarily large but finite number. If $\Delta E_W$ were set equal to infinity one would be reproducing the Dirac vacuum.

This definition of the vacuum allows transitions from the occupied negative energy states within the band to those beneath the band, thereby making the Schwinger term vanish, and the theory is consequently gauge invariant.

The vacuum can be similarly redefined in the context of QFT. With reference to Eq. (27), the creation and destruction operators in QFT obey

$$c_{ps}|0\rangle = d_{ps}|0\rangle = 0, \tag{32}$$

while new states are created by $c_{ps}^{\dagger}$ and $d_{ps}^{\dagger}$ operating on the vacuum state |0>. Solomon [12] redefines the vacuum to be $|0_R>$, as $R \to \infty$ with the following restrictions:

$$\begin{aligned} c_{ps}|0_R\rangle &= 0, \quad p, \\ d_{ps}|0_R\rangle &= 0, \quad |p| < R, \\ d_{ps}^{\dagger}|0_R\rangle &= 0, \quad |p| > R. \end{aligned} \tag{33}$$

New states are created by

$$\begin{aligned} &c_{ps}^{\dagger}|0_R\rangle, \quad p \\ &d_{ps}|0_R\rangle, \quad |p| > R \\ &d_{ps}^{\dagger}|0_R\rangle, \quad |p| < R. \end{aligned} \tag{34}$$

Graphically, these conditions may be displayed as in Fig. 2.

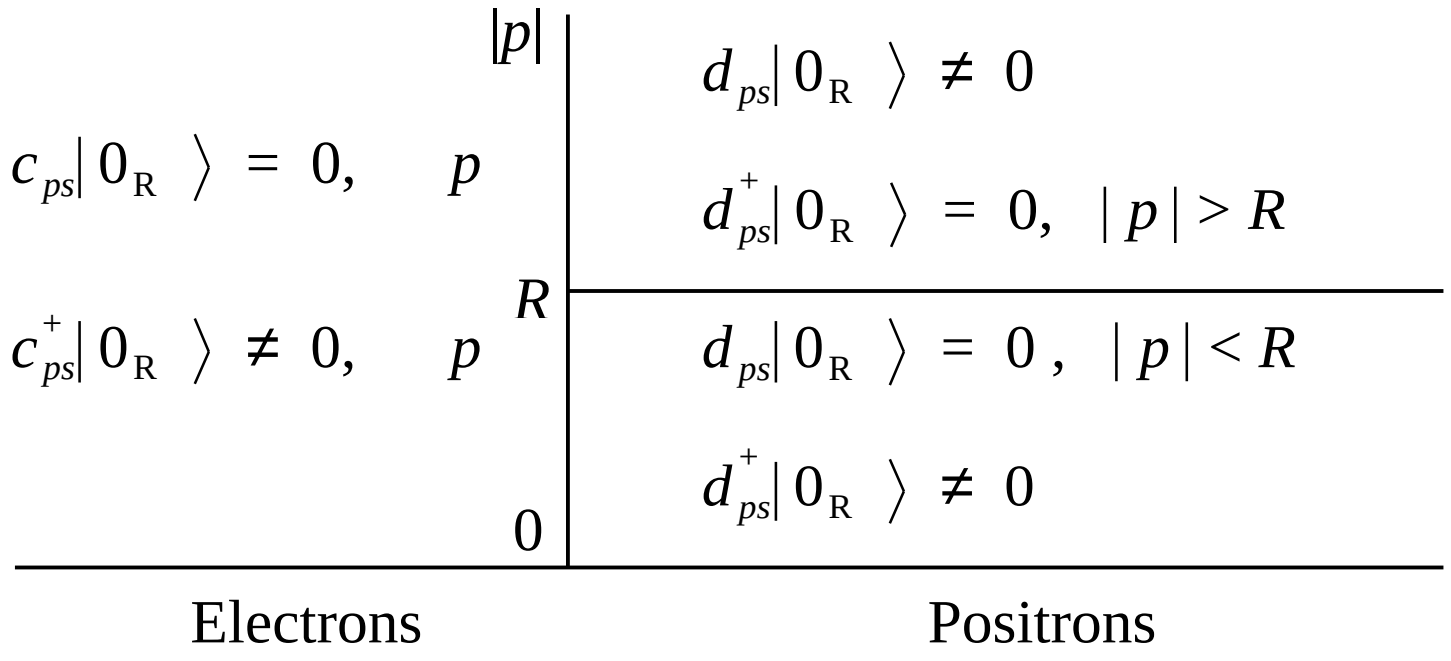

Figure 2. Solomon's [12] redefinition of the vacuum in QFT. Note that $d_{ps}|0_R>$, $|p| > R$ creates states with energy less than $|0_R>$.

As $R \to \infty$ the usual relations of Eq. (32) are recovered. Solomon shows that the Schwinger term vanishes using this redefinition of the vacuum so that the theory is gauge invariant. Again, to achieve this it was necessary to introduce negative energy states.

The key question to be addressed, with this or any redefinition of the vacuum that allows negative-energy states, is stability. First, can positive energy particles be scattered into unoccupied negative energy states? And second, is the vacuum catastrophically unstable, in the sense that the transition rate from occupied negative energy states within the band shown in Fig. 1 to those below the band, so great that the band vanishes essentially instantaneously?

With regard to the first question, the idea here is to make $\Delta E_W$ large enough to make the probability of such transitions negligibly small. The answer to the second question is still open, but Solomon [12] has determined that for a field theory of non-interacting zero-mass fermions in the presence of a classical electromagnetic field in a 2-dimensional space-time, the vacuum is stable. It remains to show that this remains the case for 4-dimensional space-time.

**Origin of the inconsistencies in QFT and Summary**

The problem of maintaining gauge invariance in QFT when interactions are present, and the difficulties with Schwinger terms discussed above, stem from the now well known fact that the underlying assumptions of QFT are inconsistent. The essence is contained in Haag's theorem [13], which is concerned with the interaction picture that forms the basis for perturbation theory. Haag's theorem is important not least for the fact that it identifies the reason regularization and renormalization are needed in QFT; that is, the underlying assumptions of relativistic QFT are inconsistent in the context of interacting systems.

In QFT, relativistic transformations between states are governed by the continuous unitary representations of the inhomogeneous group SL(2,C)—essentially the complex Poincaré group. One might anticipate that when interactions are present the unitarity condition might be violated. Indeed, Haag's theorem states, in essence, that if $\varphi_0$ and $\varphi$ are field operators defined respectively in Hilbert spaces $H_0$ and $H$, with vacua $|0>_0$ and $|0>$, and if $\varphi_0$ is a free field of mass *m*, then a unitary transformation between $\varphi_0$ and $\varphi$ exists only if $\varphi$ is also a free field of mass *m*. Another way of putting this is that if the interaction picture is well defined, it necessarily describes a free field.

The assumption that the vacuum state is the minimum energy state, invariant under a unitary transformation, is one of the fundamental assumptions of QFT. But it is now known that the physical vacuum state is not simple and must allow for spontaneous symmetry breaking and a host of other properties, so that the real vacuum bears little relation to the vacuum state of axiomatic QFT. Nevertheless, even if the latter type of vacuum is assumed, the violation of the unitarity condition in the presence of interactions opens up the possibility that the spectral condition, which limits momenta to being within or on the forward light cone, may also be violated thereby allowing negative energy states.

Of course, the way QFT gets around the formal weakness of using the interaction picture is to regularize the singular field functions that appear in the perturbation series followed by renormalization. There is nothing wrong with this approach from a pragmatic point of view, and it works exceptionally well in practice. The reason one may want to look at other approaches, such as redefining the vacuum state and the role of negative energies, is that it may lead to insights into the nature of the vacuum itself, and help resolve the outstanding cosmological constant problem, and could offer a new approach and possibly an alternative to the process of regularization and renormalization.

If it is indeed possible to use the negative energy spectrum to cancel much of the vacuum energy density of Eq. (5), there is now observational evidence pointing to what the residual vacuum energy density must be. This comes from the observational data of the Supernova Cosmology Project [14].